\documentclass[journal]{IEEEtran}

\usepackage{multirow}
\usepackage{cite}
\usepackage{amsmath,amssymb,amsfonts}
\usepackage{algorithmic}
\usepackage{graphicx}
\usepackage{textcomp}
\usepackage{xcolor}
\usepackage[flushleft]{threeparttable}
\usepackage{siunitx}
\def\BibTeX{{\rm B\kern-.05em{\sc i\kern-.025em b}\kern-.08em
    T\kern-.1667em\lower.7ex\hbox{E}\kern-.125emX}}

\begin{document}
%

\title{mustGAN: Multi-Stream Generative Adversarial Networks for MR Image Synthesis}
\author{Mahmut~Yurt,
        Salman~UH~Dar,
        Aykut~Erdem,
        Erkut~Erdem,
        and~Tolga~\c{C}ukur*,~\IEEEmembership{Senior Member,~IEEE}
\thanks{This work was supported in part by a European Molecular Biology Organization Installation Grant (IG 3028), by a TUBA GEBIP 2015 fellowship, by a BAGEP 2017 fellowship, and by a TUBITAK 1001 Research Grant (118E256). 
\textit{Asterisk indicates the corresponding author.}}
\thanks{M. Yurt, S.U.H. Dar, and T. \c{C}ukur are with the Department of Electrical and Electronics Engineering and National Magnetic Resonance Research Center, Bilkent University, Ankara, TR-06800 Turkey. T. \c{C}ukur is also with the Neuroscience Program, Sabuncu Brain Research Center, Bilkent University, Ankara, TR-06800 Turkey. (e-mail: mahmut@ee.bilkent.edu.tr; salman@ee.bilkent.edu.tr; cukur@ee.bilkent.edu.tr).}
\thanks{A. Erdem and E. Erdem are with the Department of Computer Engineering, Hacettepe University, Beytepe, 06800 Ankara (e-mail:  aykut@cs.hacettepe.edu.tr, erkut@cs.hacettepe.edu.tr)}}

\markboth{}%
{Yurt \MakeLowercase{\textit{et al.}}: Multi-Stream GANs for MR Image Synthesis}
%


\maketitle

\begin{abstract}
Multi-contrast MRI protocols increase the level of morphological information available for diagnosis. Yet, the number and quality of contrasts is limited in practice by various factors including scan time and patient motion. Synthesis of missing or corrupted contrasts can alleviate this limitation to improve clinical utility. Common approaches for multi-contrast MRI involve either one-to-one and many-to-one synthesis methods. One-to-one methods take as input a single source contrast, and they learn a latent representation sensitive to unique features of the source. Meanwhile, many-to-one methods receive multiple distinct sources, and they learn a shared latent representation more sensitive to common features across sources. For enhanced image synthesis, here we propose a multi-stream approach that aggregates information across multiple source images via a mixture of multiple one-to-one streams and a joint many-to-one stream. The shared feature maps generated in the many-to-one stream and the complementary feature maps generated in the one-to-one streams are combined with a fusion block. The location of the fusion block is adaptively modified to maximize task-specific performance. Qualitative and quantitative assessments on \textbf{$\mathrm{T_1}$}-, \textbf{$\mathrm{T_2}$}-, \textbf{$\mathrm{PD}$}-\textbf{$\mathrm{weighted}$} and \textbf{$\mathrm{FLAIR}$} images clearly demonstrate the superior performance of the proposed method compared to previous state-of-the-art one-to-one and many-to-one methods.
\end{abstract}


%
\IEEEpeerreviewmaketitle

\newcommand\tab[1][0.8cm]{\hspace*{#1}}
\newcommand\tabx[1][0.6cm]{\hspace*{#1}}
\section{Introduction}

Magnetic resonance imaging (MRI) enables a given anatomy to be imaged under different tissue contrasts by simply manipulating pulse sequences. In turn, images acquired in several distinct contrasts help better distinguish tissues and increase diagnostic information. For instance, gray and white matter can be better delineated in $\mathrm{T_1}$-weighted brain images, whereas fluids and cortical tissues can be better delineated in $\mathrm{PD}$-weighted images. Yet, multi-contrast acquisitions often prove impractical due to scan time limitations or excessive artifacts related to patient motion \cite{patient_motion,mr_artifacts}. Therefore, within-domain synthesis of missing or corrupted contrasts from other high-quality contrasts is a promising tool to improve the clinical feasibility and utility of multi-contrast MRI \cite{mr_synthesis_benefits}.

\par
Prior synthesis methods proposed for multi-contrast MRI can be grouped under two main titles depending on their input: one-to-one \cite{patch_based_one_to_one_1,patch_based_one_to_one_2,patch_based_one_to_one_4,dictionary_one_to_one_1,com_sen_mr_tissue,patch_based_one_to_one_3,loc_sens_nn_1,
dict_learning_im_synth,example_based,atlas_based_intensity,simulta_super_res,nn_one_to_one_1,nn_one_to_one_2,pgan_cgan,3D_cgan,mr_tra_seg,eagan,anonymization,cyclegan_unit} and many-to-one methods \cite{les_seg,replica,jog_reconstruct,jog_tree_encoded,multimodal,robust_multimodal_mr,mmgan,collagan,mra_synth,dar2018synergistic,diamondgan,FLAIR_MR,synthetic_flair_gan,rsnet}. In one-to-one synthesis, the goal is to generate a subject's image $y$ in a target contrast $c_T$ from the same subject's image $x$ in a source contrast $c_S$. Earlier studies have formulated one-to-one synthesis as a sparse dictionary reconstruction problem \cite{patch_based_one_to_one_2,patch_based_one_to_one_4,dictionary_one_to_one_1,com_sen_mr_tissue,dict_learning_im_synth,example_based}, where patch-based dictionaries are formed from a set of co-registered atlas image $b_S$ of $c_S$ and atlas image $b_T$ of $c_T$. Each patch in $x$ is expressed as a sparse linear combination of dictionary atoms of $b_S$, and this combination is then used for synthesizing $y$ from $b_T$ \cite{patch_based_one_to_one_2,patch_based_one_to_one_4,dictionary_one_to_one_1,com_sen_mr_tissue,dict_learning_im_synth,example_based}. For improved performance, patch-based non-linear regression using random forests \cite{patch_based_one_to_one_3} or location-sensitive neural networks \cite{loc_sens_nn_1} has been proposed for source to target mapping. To overcome limitations due to patch-based processing, later studies have proposed deep network models that process the entire source image with convolutional layers \cite{nn_one_to_one_1,nn_one_to_one_2}. One powerful method is based on the encoder-decoder architecture \cite{nn_one_to_one_2}, where latent representations of the source image are embedded via an encoder and the target image is then recovered via a decoder from these representations \cite{nn_one_to_one_2}. Recent deep network models have further incorporated an adversarial loss to better capture the high frequency details in the target image \cite{pgan_cgan,3D_cgan,mr_tra_seg,eagan,anonymization,cyclegan_unit}. An important adversarial method is pGAN \cite{pgan_cgan}, which additionally utilizes pixel-wise and perceptual losses \cite{perceptual_loss} to enhance synthesis performance. While most one-to-one synthesis methods assume spatial alignment between source and target images, adversarial networks based on cycle-consistency loss have also been proposed to mitigate this limitation \cite{pgan_cgan}. 

\par
When several source contrasts are available in a multi-contrast MRI protocol, a natural alternative is to perform many-to-one synthesis \cite{les_seg,replica,jog_reconstruct,jog_tree_encoded,multimodal,robust_multimodal_mr,mmgan,collagan,mra_synth,dar2018synergistic,diamondgan,FLAIR_MR,synthetic_flair_gan,rsnet}. In this approach, the goal is to generate the subject's image $y$ in the target contrast $c_T$ from a collection of source images $X \{x_i:i=1,2,...,K\}$ in varying contrasts $C_S \{c_{S_i}:i=1,2,...,K\}$. As in one-to-one synthesis, a common method is to perform non-linear regression using random forests \cite{replica,jog_reconstruct}. The random-forest method \cite{replica} fits a non-linear regression model in feature space to estimate intensities of the target contrast given multiple source contrasts \cite{replica}. Deep neural network methods have also been proposed for many-to-one synthesis \cite{robust_multimodal_mr,multimodal,rsnet}. In \cite{multimodal}, latent representations of multiple source contrast images are encoded through separate encoder architectures. These latent representations are then used to synthesize the target image through a joint decoder architecture \cite{multimodal}. Similar to one-to-one methods, recent studies have leveraged an adversarial loss to improve the quality of many-to-one synthesis \cite{mmgan,collagan,mra_synth,dar2018synergistic,diamondgan,FLAIR_MR,synthetic_flair_gan}. Important examples are MM-GAN \cite{mmgan} that work with spatially aligned source-target images and CollaGAN \cite{collagan} that uses cycle-consistency loss to allow for unaligned source-target images. 

\par
In general, one-to-one synthesis methods attempt to recover the target image from the latent representation of a given source image. Since these methods are optimized for a single input channel, they can sensitively learn the unique, detailed features of the given source contrast. While this sensitivity can be preferable when the images of the source and target contrast are highly correlated, it might limit synthesis performance when the two contrasts are weakly linked. On the other hand, many-to-one synthesis methods aim to recover the target image from a shared latent representation of multiple source images. These methods naturally manifest increased sensitivity for capturing features that are shared across distinct source contrasts, even when these features are weakly present in individual contrasts. Yet, a shared latent representation might also be less sensitive to complementary features that are uniquely present in a specific source contrast. When such unique information is predominantly predictive of the target image, many-to-one synthesis might perform suboptimally.

\par
Here we propose a novel method, multi-stream generative adversarial network (mustGAN), for enhanced image synthesis in multi-contrast MRI. To alleviate limitations of one-to-one and many-to-one synthesis, mustGAN leverages both shared and complementary features of multiple source images via a mixture of multiple one-to-one streams and a joint many-to-one stream. The shared feature maps generated in the many-to-one stream and the complementary feature maps generated in the one-to-one streams are later combined with a fusion block. The optimal position of the fusion for multi-contrast MRI synthesis is adaptively learned to maximize task-specific performance. A joint network is then trained to generate the target image from the fused feature maps. Comprehensive assessments are performed on multi-contrast MR images ($\mathrm{T_1}$-, $\mathrm{T_2}$-, $\mathrm{PD}$-weighted and $\mathrm{FLAIR}$ images) from healthy subjects and high/low grade glioma patients. The proposed method yields both quantitavely and qualitatively higher performance in multi-contrast MR image synthesis as compared to state-of-the-art one-to-one and many-to-one methods.

\section{Theory}

\subsection{Generative Adversarial Networks}
A GAN consists of a pair of competing networks; a generator ($G$) and a discriminator ($D$) \cite{gan}. In an image synthesis task, $G$ maps a random noise vector $z$ to an output image $y$ from a target distribution $p(y)$, $G:z\rightarrow y$, and $D$ estimates the probability that a sample $s$ is drawn from $p(y)$, $D:s$. While $G$ is trained to synthesize fake images that are indistinguishable from real images, $D$ is trained to discriminate between real and generated images \cite{gan}. This can be formulated as a minimax game based on an adversarial loss function $L_{GAN}$.
\begin{equation}
\begin{aligned}
\min_{G} \max_{D} L_{GAN} = &\min_{G} \max_{D} \big(E_y [logD(y)]\\& +E_z[log(1-D(G(z)))] \big)
\end{aligned}
\end{equation}
where $E$ denotes expectation. To improve stability, negative log-likelihood in $L_{GAN}$ is typically replaced by a squared loss function \cite{lsgan}:
\begin{equation}
L_{GAN}=-E_y [(D(y)-1)^2 ]-E_z [D(G(z))^2 ]
\end{equation}

\subsection{Conditional Generative Adversarial Networks}

Recent studies on image-to-image translation have demonstrated that conditional GANs (condGANs) are highly effective in mapping between statistically-dependent source and target images \cite{condgans}, i.e., when these images manifest the same underlying scene in distinct domains. To capture this dependency, condGANs take as input the source image $x$ as prior information \cite{condgans}. The adversarial loss is then expressed as follows:
\begin{equation}
L_{condGAN}=-E_{xy} [(D(x,y)-1)^2 ]-E_{x} [D(G(x))^2 ]
\end{equation}
When source and target images are spatially registered, a pixel-wise loss can be added between the ground truth and generated images \cite{pix2pix}:
\begin{equation}
L_{pixel-wise}=E_{xy} [\Big|\Big|y-G(x)\Big|\Big|_1 ]
\end{equation}
The joint loss function then becomes:
\begin{equation}
\label{eqn:eq_1}
\begin{aligned}
L_{condGAN}=&-E_{xy} [(D(x,y)-1)^2 ]-E_{x} [D(x,G(x))^2 ]\\&+E_{xy} [\Big|\Big|y-G(x)\Big|\Big|_1 ]
\end{aligned}
\end{equation}

In a previous study, we have demonstrated that condGAN-based architectures yield state-of-the-art performance for one-to-one MR image synthesis (e.g. $\mathrm{T_1}\rightarrow \mathrm{T_2}$ and $\mathrm{T_2}\rightarrow \mathrm{T_1}$) \cite{pgan_cgan}. Yet, a multitude of different contrasts are often collected in an MR exam. When multiple source images are available, many-to-one condGAN models may offer improved performance. Given $K$ source contrast images denoted as $X\{ x_m:m=1,2,...,K\}$, a many-to-one condGAN model is formulated as:
\begin{equation}
\begin{aligned}
L_{condGAN}=&-E_{x_1x_2...x_Ky}[(D(x_1,x_2,...,x_K,y)-1)^2]\\&-E_{x_1x_2...x_K}[D(x_1,x_2,...,x_K,\\ & \tab \tab \tabx G(x_1,x_2,...,x_K))^2]\\&+E_{x_1x_2...x_Ky}[\Big|\Big|y-G(x_1,x_2,...,x_K)\Big|\Big|_1 ]
\end{aligned}
\end{equation}
Note that this formulation corresponds to a single-stream many-to-one network that concatenates multiple source contrast images at the input level. 

\section{Methods}

\subsection{Multi-Stream GAN Model}\label{AA}
\begin{figure*}[htbp]
\centerline{\includegraphics[width=0.85\textwidth]{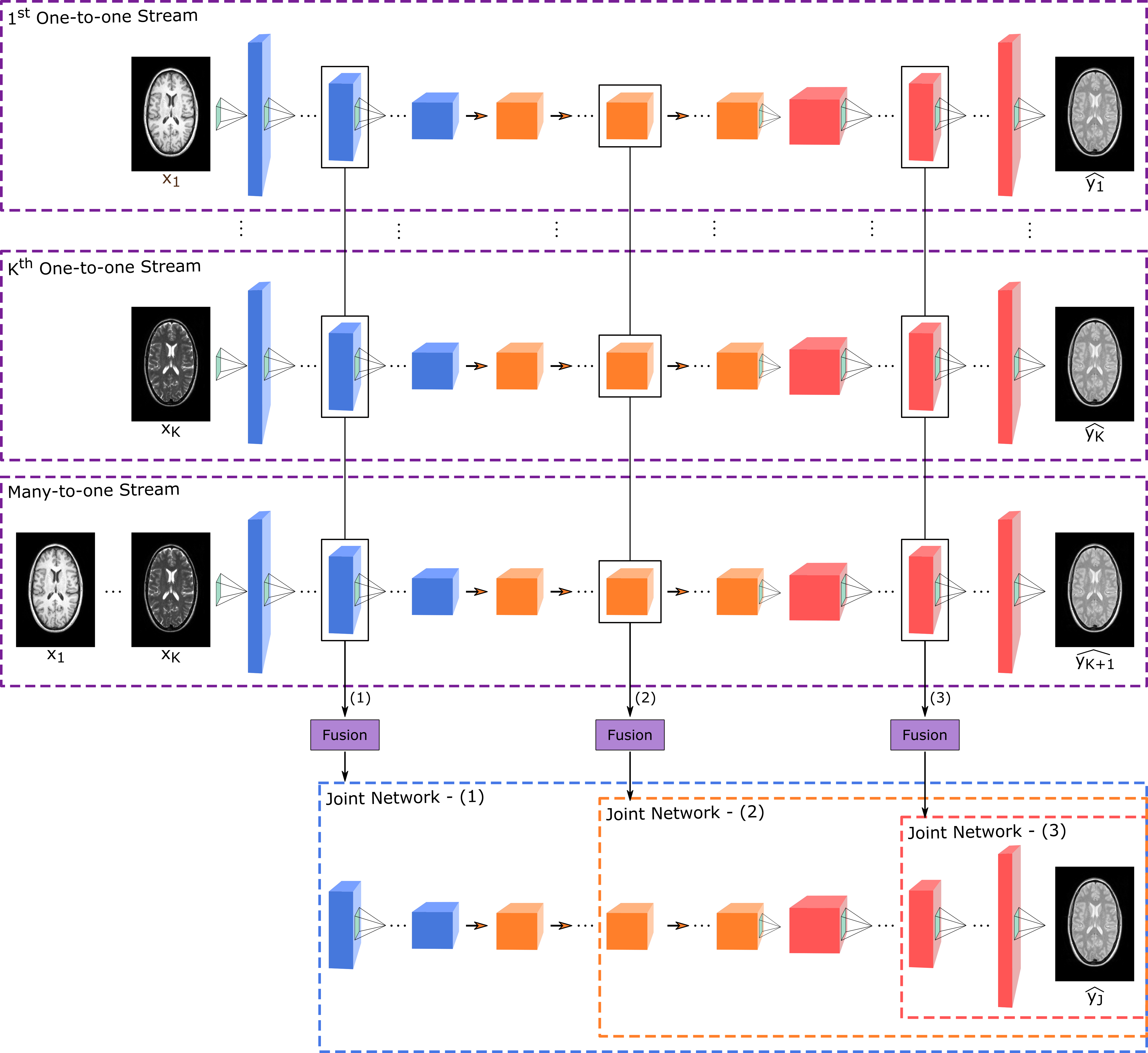}}
\caption{The generator (G) in mustGAN consists of $K$ one-to-one streams and a many-to-one stream, followed by an adaptively positioned fusion block, and a joint network for finaly recovery. One-to-one streams generate the unique feature maps of each source image independently, whereas the many-to-one stream generates the shared feature map across source images. The fusion block fuses the feature maps generated in the fusion layer by concatenation. Lastly, the joint network synthesizes the target image from these fused feature maps. Note that the architecture of the joint network varies depending on the position of the fusion that is categorized under three titles: early fusion (1), intermediate fusion (2) and late fusion (3).}
\label{fig:fig_1}
\end{figure*}
Here we propose a multi-stream GAN architecture (mustGAN) that leverages information from multiple source contrasts by adaptively combining one-to-one and many-to-one streams (Fig. \ref{fig:fig_1}). To synthesize the target contrast $y$, mustGAN receives a collection of source images denoted as $X \{x_m:m=1,2,...,K\}$. First, mustGAN learns $K$ independent one-to-one streams, where each stream is a condGAN model trained to generate the target image from a distinct source image. Second, mustGAN learns a single many-to-one stream -again a condGAN model- that is trained to generate the target image from all source images concatenated at the input level. mustGAN then fuses the unique feature maps generated in one-to-one streams and the shared feature map generated in the many-to-one stream. The position of the fusion block is also learned to optimize task-specific performance. Finally, mustGAN trains a joint network that synthesizes the target image given the fused feature maps. The architecture of this joint network varies depending on the position of the fusion block.
\vskip 0.075in
\subsubsection{One-to-One Streams}
The proposed architecture contains $K$ separate one-to-one streams, where the $m^{th}$ stream synthesizes $y$ from the source contrast $x_m$ via a generator $G_m$ and a discriminator $D_m$. $G_m$ consists of three sub-networks: an encoder $(E_m)$ with $n_E$ convolutional layers, a residual network $(R_m)$ with $n_R$ ResNet blocks and a decoder $(D_m)$ with $n_D$ convolutional layers. $G_m$ is expressed as: 
\begin{equation}
\widehat{y_m}=G_m(x_m)=D_m(R_m(E_m(x_m)))
\end{equation}
where $\widehat{y_m}$ denotes the predicted target image. Meanwhile, $D_m$ is a convolutional network $(C_m)$ with $n_C$ layers:
\begin{equation}
D_m(x_m,s)=C_m(x_m,s)
\end{equation}
where $s$ is either $G_m(x_m)$ or $y$. To train $G_m$ and $D_m$, adversarial and pixel-wise losses are utilized:
\begin{equation}
\begin{aligned}
L_m=&-E_{x_my}[(D_m(x_m,y)-1)^2]\\&-E_{x_m}[D_m(x_m,G_m(x_m))^2 ]\\&+E_{x_my}[\Big|\Big|y-G_m(x_m)\Big|\Big|_1]
\end{aligned}
\end{equation}
$G_m$ aims to map $x_m$ to $y$, and $D_m$ aims to discriminate between $\widehat{y_m}$ and $y$. 
\vskip 0.15in
\subsubsection{Many-to-One Stream}
mustGAN contains a $(K+1)^{th}$ stream that performs many-to-one synthesis given all source images. This generator $G_{K+1}$ again consists of an encoder ($E_{K+1}$) with $n_E$ convolutional layers, a residual network ($R_{K+1}$) with $n_R$ ResNet blocks and a decoder ($D_{K+1}$ ) with $n_D$ convolutional layers:
\begin{equation}
\begin{aligned}
\widehat{y_{K+1}}&=G_{K+1}(x_1,x_2,...,x_K)\\&=D_{K+1}(R_{K+1}(E_{K+1} (x_1,x_2,...,x_K))) 
\end{aligned}
\end{equation}
The discriminator $D_{K+1}$ containing $n_C$ convolutional layers is formulated as:
\begin{equation}
D_{K+1}(x_1,x_2,...,x_K,s)=C_{K+1}(x_1,x_2,...,x_K,s)
\end{equation}
where $s$ is either $G_{K+1}(x_1,x_2,...,x_K)$ or $y$. The loss function for the $(K+1)^{th}$ stream is given as:
\begin{equation}
\begin{aligned}
    L_{K+1}=&-E_{x_1x_2...x_Ky}[(D_{K+1}(x_1,x_2,...,x_K,y)-1)^2]\\&-E_{x_1x_2...x_K}[D_{K+1}(x_1,x_2,...,x_K,
    \\ &  \tab \tab \tabx G_{K+1}(x_1,x_2,...,x_K))^2]\\&+E_{x_1x_2...x_Ky}[\Big|\Big|y-G_{K+1}(x_1,x_2,...,x_K)\Big|\Big|_1]
\end{aligned}
\end{equation}
$G_{K+1}$ effectively learns to predict $y$ given $x_1,x_2,...,x_K$ concatenated at the input level, and $D_{K+1}$ learns to discriminate between $\widehat{y_{K+1}}$ and $y$. 
\vskip 0.15in
\subsubsection{Joint Network}
Once the $K+1$ streams are trained, source images are propageted separately through the streams up to the fusion block ($F$) at the $i^{th}$ layer. $F$ concatenates the feature maps generated at the $i^{th}$ layer of the one-to-one and many-to-one streams. A joint network ($J$) is then trained to recover the target image from the fused feature maps. The precise architecture of $J$ varies depending on the position of $F$, considered in three types here: early, intermediate, and late fusion. 
\paragraph{Early Fusion}
Early fusion occurs when $F$ is within the encoder (i.e. $0<i< n_E$). The feature maps generated by the $m^{th}$ one-to-one stream ($g_m^i$) and by the many-to-one stream ($g_{K+1}^i$) at the $i^{th}$ layer are formulated as:
\begin{equation*}
\begin{aligned}
&g_m^i=E_m(x_m|i)
\\ &g_{K+1}^i=E_{K+1}(x_1,x_2,...,x_K|i)
\end{aligned}
\end{equation*}
These feature maps are concatenated by $F$ yielding the fused feature maps ($g_F^i$): 
\begin{equation}
    g_F^i=F(g_1^i,g_2^i,...,g_K^i,g_{K+1}^i)
\end{equation}
$J$ receives as input these fused maps to recover the target image. Thus, architecture of $J$ for early fusion is as follows:
\begin{equation}
\widehat{y}=J(g_F^i)=D_J(R_J(E_J(g_F^i|i)))
\end{equation}
\paragraph{Intermediate Fusion}
Intermediate fusion occurs when $F$ is within the residual block (i.e. $n_E\leq i<n_E+n_R$). In this case, the feature maps generated by the $m^{th}$ one-to-one stream ($g_m^i$) and the many-to-one stream ($g_{K+1}^i$) are formulated as:
\begin{equation*}
\begin{aligned}
&g_m^i=R_m(E_m(x_m)|i)
\\ &g_{K+1}^i=R_{K+1}(E_{K+1}(x_1,x_2,...,x_K)|i)
\end{aligned}
\end{equation*}
The fused feature maps ($g_F^i$) are then: 
\begin{equation}
    g_F^i=F(g_1^i,g_2^i,...,g_K^i,g_{K+1}^i)
\end{equation}
$J$ again receives as input the fused maps to recover the target image. Architecture of $J$ for intermediate fusion is as follows:
\begin{equation}
\widehat{y}=J(g_F^i)=D_J(R_J(g_F^i|i))
\end{equation}
\paragraph{Late Fusion}
Late fusion occurs when $F$ is within the decoder (i.e. $n_E+n_R\leq i< n_E+n_R+n_D$). The feature maps by the $m^{th}$ one-to-one stream ($g_m^i$) and by the many-to-one stream at the $i^{th}$ layer ($g_{K+1}^i$) are given as:
\begin{equation*}
\begin{aligned}
&g_m^i=D_m(R_m(E_m(x_m))|i)
\\ &g_{K+1}^i=D_{K+1}(R_{K+1}(E_{K+1}(x_1,x_2,...,x_K))|i)
\end{aligned}
\end{equation*}
In turn, the fused feature maps ($g_F^i$) are: 
\begin{equation}
    g_F^i=F(g_1^i,g_2^i,...,g_K^i,g_{K+1}^i)
\end{equation}
$J$ receives as input the fused maps to recover the target image, yielding the following architecture for late fusion:
\begin{equation}
\widehat{y}=J(g_F^i)=D_J(g_F^i|i)
\end{equation}

$J$ is also trained in an adversarial setup independent from $i$ together with a fixed discriminator $D_J$, which is a convolutional network ($C$) with $n_C$ layers:
\begin{equation}
    D_J(x_1,x_2,…,x_K,s)=C_J(x_1,x_2,…,x_K,s)
\end{equation}
where s is either $J(g_F^i)$ or $y$. To train $J$ and $D_J$, a loss function consisting of an adversarial loss and pixel-wise $L1$ loss is utilized:
\begin{equation}
\begin{aligned}
    L_J=&-E_{x_1x_2...x_K,y}[logD_J(x_1,x_2,...,x_K,y)]\\&-E_{x_1x_2...x_K}[log(1-D_J(x_1,x_2
    ...,x_K,J(g_F^i))) ]\\&+E_{x_1x_2...x_Ky}[\Big|\Big|y-J(g_F^i)\Big|\Big|_1]
    \end{aligned}
\end{equation}

\subsection{Network Architecture}\label{AA}
The $K$ one-to-one streams and the many-to-one stream had identical ($G$) and discriminator ($D$) architectures adopted from\cite{perceptual_loss} and \cite{cycleGAN}. $G$ consisted of an encoder ($E$) of $3$ convolutional layers, a residual network ($R$) of $9$ ResNet blocks and a decoder ($D$) of $3$ convolutional layers. $D$ consisted of a convolutional network ($C$) of $5$ layers. The number of input channels was $1$ for $G_m$, $2$ for $D_m$, $K$ for $G_{K+1}$, and $K+1$ for $D_{K+1}$.
\par
The architecture of the joint network ($J$) was adaptively modified based on the position of the fusion block ($i$). When $0<i<3$ (i.e. early fusion), $J$ consisted of $3-i$ convolutional layers, $9$ ResNet blocks and $3$ convolutional layers connected in series. When $3\leq i<12$ (i.e. intermediate fusion), $J$ consisted of $12-i$ ResNet blocks and $3$ convolutional layers. When $12\leq i<15$ (i.e. late fusion), $J$ consisted of $15-i$ convolutional layers. The corresponding discriminator $D_J$ consisted of a convolutional network ($C_J$) of $5$ layers. The number of input channels was $K+1$ for $D_J$, and variable for $J$ depending on $i$.

\subsection{Datasets}\label{AA}
Demonstrations were performed on two separate neuroimaging datasets: the IXI dataset (http://brain-development.org/ixi-dataset/) that contained data from healthy subjects and the ISLES dataset \cite{isles} that contained data from high/low grade glioma patients. Data normalization was performed to provide comparable voxel intensities across subjects. To do this, the maximum intensity of each brain volume was normalized to $1$ within individual subjects and separately for each MR contrast.

\paragraph{IXI Dataset}
$\mathrm{T_1}$-, $\mathrm{T_2}$- and $\mathrm{PD}$-weighted images from $53$ subjects were used, where $25$ were reserved for training, $10$ were reserved for validation and $18$ were reserved for testing. Subject selection was performed sequentially. Approximately $100$ axial cross sections that contained artifact-free brain tissue were manually selected from each subject. The images were acquired with the following parameters. $\mathrm{T_1}$-weighted images: $\mathrm{TE}=4.603ms$, $\mathrm{TR}=9.813ms$, $\mathrm{flip}$ $\mathrm{angle}=\ang{8}$, $\mathrm{voxel}$ $\mathrm{dimensions}=0.94\times 0.94\times 1.2 mm^3$. $\mathrm{T_2}$-weighted images: $\mathrm{TE}=100ms$, $\mathrm{TR}=8178.34ms$, $\mathrm{flip}$ $\mathrm{angle}=\ang{90}$, $\mathrm{voxel}$ $\mathrm{dimensions}=0.94\times 0.94\times 1.2mm^3$. $\mathrm{PD}$-weighted images: $\mathrm{TE}=8ms$, $\mathrm{TR}=8178.34ms$, $\mathrm{flip}$ $\mathrm{angle}=\ang{90}$, $\mathrm{voxel}$ $\mathrm{dimensions}=0.94\times 0.94\times 1.2mm^3$. Note that images of separate contrasts were unregistered in this dataset. Hence, $\mathrm{T_2}$- and $\mathrm{PD}$-weighted images were registered onto $\mathrm{T_1}$-weighted images by rigid transformation based on mutual-information. Registration was performed via FSL.

\paragraph{ISLES Dataset}
$\mathrm{T_1}$-, $\mathrm{T_2}$-weighted and $\mathrm{FLAIR}$ images from $56$ subjects were used, where $25$ were reserved for training, $10$ were reserved for validation and $21$ were reserved for testing. Subject selection was performed sequentially. Approximately $100$ axial cross sections that contained artifact-free brain tissue were manually selected from each subject. The ISLES dataset comprised images collected under a heterogeneous set of scanning parameters \cite{isles}. Since $\mathrm{T_1}$- and $\mathrm{T_2}$-weighted images were already aligned to $\mathrm{FLAIR}$ images\cite{isles}, no other registration was performed.

\subsection{Network Training}\label{AA}
The network training procedure for mustGAN comprises two sequential phases: the individual training of the one-to-one and many-to-one streams, and the training of the joint network following fusion. For the first phase, we adopted hyperparameter selection from a previous study \cite{pgan_cgan}, where successful one-to-one image synthesis was demonstrated in multi-contrast MRI via conditional GAN models. The streams were trained for $100$ epochs via the Adam optimizer\cite{adam}, where the learning rate was set to $2\times 10^{-4}$ in the first $50$ epochs, and was linearly decayed from $2\times 10^{-4}$ to $0$ in the last $50$ epochs. During the training, the decay rates of the first moment $\beta_1$ and the second moment $\beta_2$ of gradient estimates were set to $0.5$ and $0.999$, respectively. Relative weighting of the pixel-wise loss to adversarial loss was selected as $100$. 
\par
For the second phase, we performed analyses to determine the optimal position of the fusion block for each synthesis task. Since the complexity of the joint network also depends on the position of the fusion block, we reasoned that the required number of epochs for convergence should also be optimized. Therefore, we performed one-fold cross validation with grid search for the fusion block position and number of epochs. To do this, multiple joint network architectures were trained for varying number of epoch values (i.e. $5:5:100$) and fusion block positions (i.e. $1:1:14$). Remaining hyperparameters were again adopted from \cite{pgan_cgan}. During the training, the Adam optimizer was employed, where the decay rates of the first moment $\beta_1$ and the second moment $\beta_2$ of gradient estimates were set to $0.5$ and $0.999$, respectively. Relative weighting of the pixel-wise loss to adversarial loss was selected as $100$. For models trained for fewer than $50$ epochs the learning rate was set to $2\times 10^{-4}$, and for models trained for more than $50$ epochs the learning rate was set to $2\times 10^{-4}$ in the first $50$ epochs and decreased by $4\times 10^{-5}$ in each remaining epoch. Based on performance evaluations on the validation set, task-specific values for the position of the fusion block and the number of epochs denoted as (fusion block position, number of epochs) were determined to be $(12, 40)$ for $\mathrm{T_1}$-weighted, $(14,15)$ for $\mathrm{T_2}$-weighted and $(12,20)$ for $\mathrm{PD}$-weighted image synthesis in the IXI dataset and $(8,50)$ for $\mathrm{T_1}$-weighted, $(6,70)$ for $\mathrm{T_2}$-weighted and $(6,10)$ for $\mathrm{FLAIR}$ image synthesis in the ISLES dataset. Note that while training the joint network, the neural network layers in the one-to-one and many-to-one streams prior to the fusion block were also fine-tuned. To do this, the Adam optimizer was employed with half the learning rate of the joint network. The decay rates of the first moment $\beta_1$ and the second moment $\beta_2$  of gradient estimates were again set to $0.5$ and $0.999$, respectively.

\subsection{Competing Methods}
To comparatively evaluate the performance of mustGAN, two variants of a state-of-the-art adversarial method for multi-contrast MRI synthesis, pGAN, were implemented. As the first variant (pGAN$_{\mathrm{one}}$), the original pGAN architecture was implemented by utilizing the adversarial and pixel-wise losses. Multiple pGAN$_{\mathrm{one}}$ models were trained, where each receives a distinct source contrast to generate the target contrast. As the second variant (pGAN$_{\mathrm{many}}$), the number of input channels of the original pGAN architecture was set to the number of available source contrast images. Again adversarial and pixel-wise losses were utilized. Here pGAN$_{\mathrm{one}}$ and pGAN$_{\mathrm{many}}$ correspond to the independently trained one-to-one and many-to-one streams in mustGAN, respectively. Therefore, pGAN$_{\mathrm{one}}$ and pGAN$_{\mathrm{many}}$ were trained with the same hyperparameter set used for the one-to-one and many-to-one streams.

\subsection{Experiments}
Two public multi-contrast MRI datasets (IXI and ISLES) were used to evaluate the performance of the proposed method. In the IXI dataset, for pGAN$_{\mathrm{one}}$, $6$ distinct cases for direction of synthesis were examined ($\mathrm{T_2}\rightarrow \mathrm{T_1}$; $\mathrm{PD}\rightarrow \mathrm{T_1}$; $\mathrm{T_1}\rightarrow \mathrm{T_2}$; $\mathrm{PD}\rightarrow \mathrm{T_2}$; $\mathrm{T_1}\rightarrow \mathrm{PD}$; $\mathrm{T_2}\rightarrow \mathrm{PD}$). For pGAN$_{\mathrm{many}}$, $3$ distinct cases for direction of synthesis were examined ($\mathrm{T_2}, \mathrm{PD}\rightarrow \mathrm{T_1}$; $\mathrm{T_1}, \mathrm{PD}\rightarrow \mathrm{T_2}$; $\mathrm{T_1}, \mathrm{T_2}\rightarrow \mathrm{PD}$). For mustGAN, $3$ distinct cases for direction of synthesis were examined ($\mathrm{T_2}, \mathrm{PD}\rightarrow \mathrm{T_1}$; $\mathrm{T_1}, \mathrm{PD}\rightarrow \mathrm{T_2}$; $\mathrm{T_1}, \mathrm{T_2}\rightarrow \mathrm{PD}$), along with $14$ distinct levels for the position of the fusion block ($1:1:14$). Overall $6$ independent pGAN$_{\mathrm{one}}$, $3$ independent pGAN$_{\mathrm{many}}$ and $42$ independent mustGAN models were trained. 
\par
In the ISLES dataset, for pGAN$_{\mathrm{one}}$, $6$ distinct cases for direction of synthesis were examined ($\mathrm{T_2}\rightarrow \mathrm{T_1}$; $\mathrm{FLAIR}\rightarrow \mathrm{T_1}$; $\mathrm{T_1}\rightarrow \mathrm{T_2}$; $\mathrm{FLAIR}\rightarrow \mathrm{T_2}$; $\mathrm{T_1}\rightarrow \mathrm{FLAIR}$; $\mathrm{T_2}\rightarrow \mathrm{FLAIR}$). For pGAN$_{\mathrm{many}}$, $3$ distinct cases for direction of synthesis were examined ($\mathrm{T_2}, \mathrm{FLAIR}\rightarrow \mathrm{T_1}$; $\mathrm{T_1}, \mathrm{FLAIR}\rightarrow \mathrm{T_2}$; $\mathrm{T_1}, \mathrm{T_2}\rightarrow \mathrm{FLAIR}$). For mustGAN, $3$ distinct cases for direction of synthesis were examined ($\mathrm{T_2}, \mathrm{FLAIR}\rightarrow \mathrm{T_1}$; $\mathrm{T_1}, \mathrm{FLAIR}\rightarrow \mathrm{T_2}$; $\mathrm{T_1}, \mathrm{T_2}\rightarrow \mathrm{FLAIR}$), along with $14$ levels for the fusion block ($1:1:14$). Overall, $6$ independent pGAN$_{\mathrm{one}}$, $3$ independent pGAN$_{\mathrm{many}}$ and $42$ independent mustGAN models were trained. 
\par
All methods were trained and tested on the same set of data samples. To quantitatively assess synthesis performance, synthesizes target images were compared with the ground truth images based on structural similarity index (SSIM) \cite{ssim} and peak signal-to-noise ratio (PSNR). Prior to measurement, voxel intensities of synthesized and ground truth images were normalized to a maximum of $1$. To assess significance of differences between competing methods, non-parametric Wilcoxon signed-rank test was employed. All training and evaluation procedures were run on NVIDIA Titan X Pascal and Xp GPUs. Implementations of mustGAN, pGAN$_{\mathrm{one}}$ and pGAN$_{\mathrm{many}}$ were performed via the PyTorch framework in Python.

\section{Results}
\begin{figure*}[htbp]
\centerline{\includegraphics[width=1\textwidth]{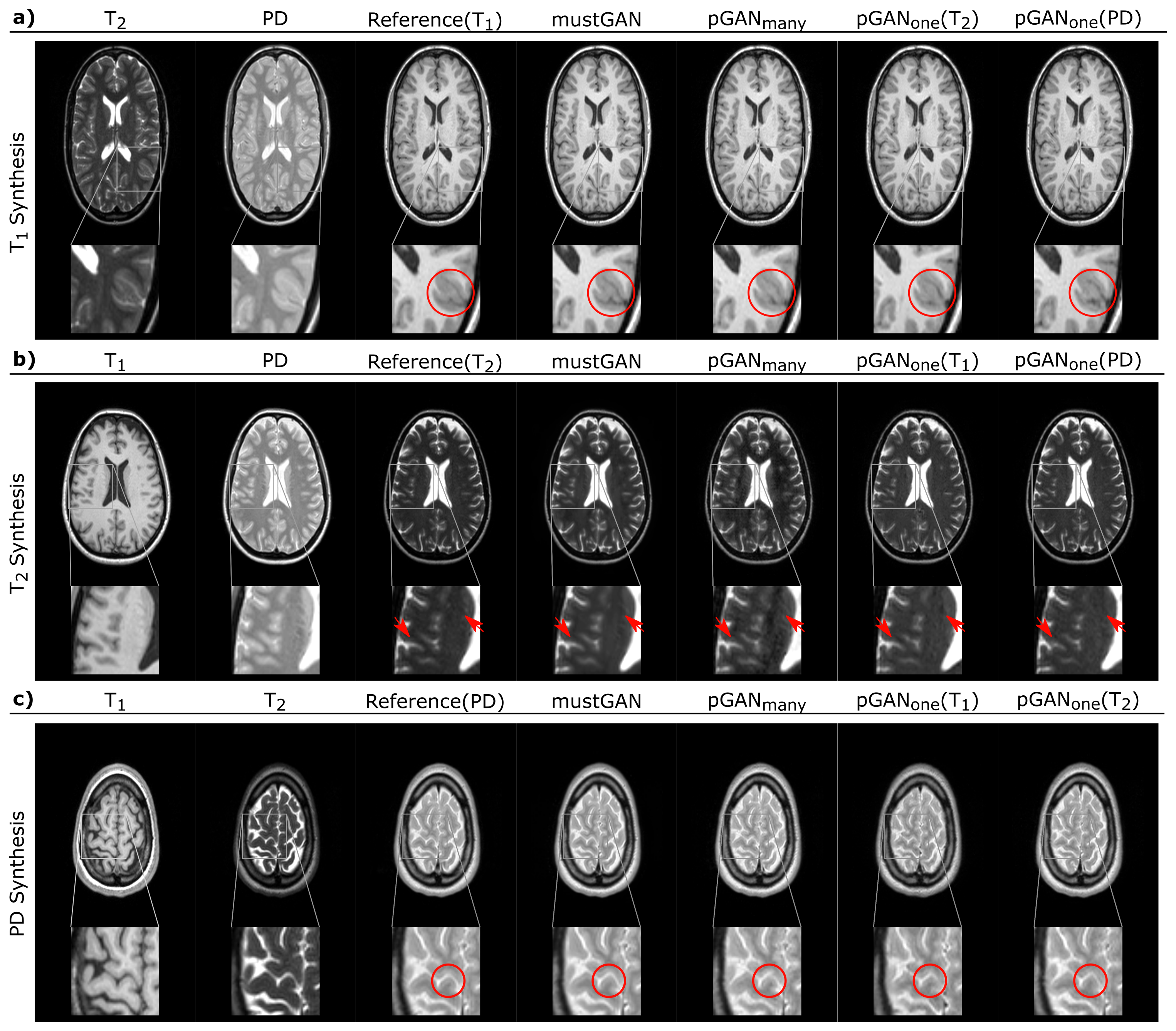}}
\caption{The proposed method was demonstrated on healthy subjects from the IXI dataset for three synthesis tasks: a) $\mathrm{T_1}$-weighted image synthesis from $\mathrm{T_2}$- and $\mathrm{PD}$-weighted images, b) $\mathrm{T_2}$-weighted image synthesis from $\mathrm{T_1}$- and $\mathrm{PD}$-weighted images, c) $\mathrm{PD}$-weighted image synthesis from $\mathrm{T_1}$- and $\mathrm{T_2}$-weighted images. Synthesized images from mustGAN, pGAN$_{\mathrm{many}}$ and pGAN$_{\mathrm{one}}$ are shown along with the source images (first two columns) and the ground truth target image (third column). Due to synergistic use of information captured by one-to-one and many-to-one streams, mustGAN improves synthesis accuracy in many regions that are recovered suboptimally in competing methods (marked with arrows or circles in zoom-in displays). Overall, mustGAN yields less noisy depiction of white-matter tissues, and sharper depiction of gray-matter tissue boundaries.}
\label{fig}
\end{figure*}

\begin{figure*}[htbp]
\centerline{\includegraphics[width=1\textwidth]{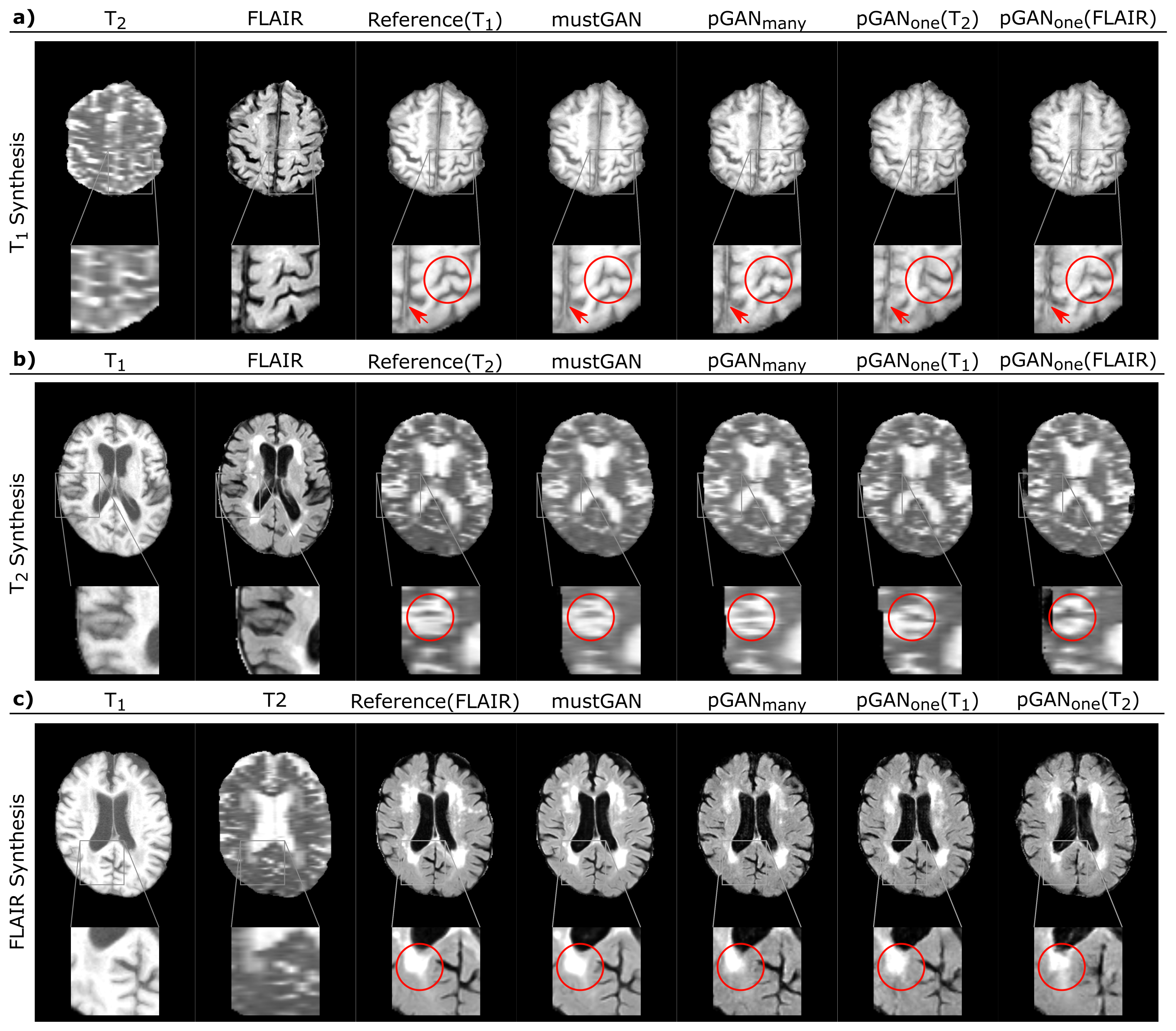}}
\caption{The proposed method was demonstrated on high/low grade glioma patients from the ISLES dataset for three synthesis tasks: a) $\mathrm{T_1}$-weighted image synthesis from $\mathrm{T_2}$-weighted and $\mathrm{FLAIR}$ images, b) $\mathrm{T_2}$-weighted image synthesis from $\mathrm{T_1}$-weighted and $\mathrm{FLAIR}$ images, c) $\mathrm{FLAIR}$ image synthesis from $\mathrm{T_1}$- and $\mathrm{T_2}$-weighted images. Synthesized images from mustGAN, pGAN$_{\mathrm{many}}$ and pGAN$_{\mathrm{one}}$ are shown along with the source images (first two columns) and the ground truth target image (third column). Due to synergistic use of information captured by one-to-one and many-to-one streams, mustGAN improves synthesis accuracy in many regions that are recovered suboptimally in competing methods (marked with arrows or circles in zoom-in displays). Overall, mustGAN yields less noisy depiction of white-matter tissues, and sharpeR depiction of gray-matter tissue boundaries.}
\label{fig}
\end{figure*}
\subsection{Task-Specific Fusion Across Multiple Streams}
To optimize the mustGAN model for specific tasks, we performed experiments to determine the optimal position of the fusion block in the architecture. Multiple mustGAN models were trained while varying the layer of fusion in the range $[1$ $14]$. Model performances were evaluated on the validation set based on PSNR measurements. Experiments were conducted separately on the IXI and ISLES datasets. In the IXI dataset, we considered three synthesis tasks: $\mathrm{T_2}, \mathrm{PD}\rightarrow \mathrm{T_1}$; $\mathrm{T_1}, \mathrm{PD}\rightarrow \mathrm{T_2}$; $\mathrm{T_1}, \mathrm{T_2}\rightarrow \mathrm{PD}$. Performance as a function of fusion layer is plotted in Supp. Fig. 1a-c for $\mathrm{T_1}$-weighted, $\mathrm{T_2}$-weighted and $\mathrm{PD}$-weighted image synthesis, respectively. Across all synthesis tasks in the IXI dataset, mustGAN models performing late fusion mostly yielded enhanced synthesis performance. Particularly, the optimal position of the fusion block is determined to be the  $12^{th}$ layer for $\mathrm{T_1}$ synthesis, the $14^{th}$ layer for $\mathrm{T_2}$ synthesis, and the $12^{th}$ layer for $\mathrm{PD}$ synthesis. Furthermore, proper selection of the fusion layer has a noticable effect on model performance, where the PSNR difference between highest and lowest performing models is $0.794$ $\mathrm{dB}$ for $\mathrm{T_1}$ synthesis, $1.057$ $\mathrm{dB}$ for $\mathrm{T_2}$ synthesis, and $0.655$ $\mathrm{dB}$ for $\mathrm{PD}$ synthesis.

\par
In the ISLES dataset, we considered three different synthesis tasks: $\mathrm{T_2}, \mathrm{FLAIR}\rightarrow \mathrm{T_1}$; $\mathrm{T_1}, \mathrm{FLAIR}\rightarrow \mathrm{T_2}$; $\mathrm{T_1}, \mathrm{T_2}\rightarrow \mathrm{FLAIR}$. Performance as a function of fusion layer is plotted in Supp. Fig. 2a-c for $\mathrm{T_1}$-weighted, $\mathrm{T_2}$-weighted and $\mathrm{FLAIR}$ image synthesis, respectively. Across all synthesis tasks in the ISLES dataset, mustGAN models performing intermediate fusion mostly yielded enhanced synthesis performance. Particularly, the optimal position of the fusion block is determined to be the  $8^{th}$ layer for $\mathrm{T_1}$ synthesis, the $6^{th}$ layer for $\mathrm{T_2}$ synthesis, and the $6^{th}$ layer for $\mathrm{FLAIR}$ synthesis. Again, proper selection of the fusion layer has a noticable effect on model performance, where the PSNR difference between highest and lowest performing models is $0.621$ $\mathrm{dB}$ for $\mathrm{T_1}$ synthesis, $0.499$ $\mathrm{dB}$ for $\mathrm{T_2}$ synthesis, and $0.623$ $\mathrm{dB}$ for $\mathrm{FLAIR}$ synthesis. These task-specific fusion layers identified on the IXI and ISLES datasets were utilized in all evaluations thereafter.

\par
Here we observed that the optimal position of the fusion block varies between the datasets. In IXI, synthesis quality is enhanced by performing the fusion within the decoder, where the fused feature maps have larger width and height and so they reflect a high-resolution representation. On the other hand, in ISLES, synthesis quality is enhanced by performing the fusion within the residual block, where the fused feature maps have smaller size, reflecting a relatively low-resolution representation. Note that the IXI dataset mainly contains high-quality, high-SNR images, so fusion at the decoder might help better recover fine structural detaills. In contrast, the ISLES dataset mostly contains limited resolution images, so fusing at the residual block might help better recover global structural information.

\subsection{Demonstrations Against Competing Methods}
Next, we comparatively evaluated the performance of mustGAN against several state-of-the-art one-to-one and many-to-one methods (pGAN$_{\mathrm{one}}$ and pGAN$_{\mathrm{many}}$). In the IXI dataset, we considered: $\mathrm{T_2}, \mathrm{PD}\rightarrow \mathrm{T_1}$; $\mathrm{T_1}, \mathrm{PD}\rightarrow \mathrm{T_2}$; $\mathrm{T_1}, \mathrm{T_2}\rightarrow \mathrm{PD}$. Pairwise comparisons across cross-sections in the test set between the proposed and competing methods are displayed in Supp. Figs. 3-5 for $\mathrm{T_1}$, $\mathrm{T_2}$ and $\mathrm{PD}$ synthesis, respectively. On average across tasks, pGAN$_\mathrm{one}$ yields higher PSNR than pGAN$_\mathrm{many}$ in 18.00\% of test samples, suggesting that unique features of source contrasts can be critical for successful synthesis. Meanwhile, mustGAN reduces this proportion to merely 7.81\% by pooling information from one-to-one and many-to-one streams. Table I lists the PSNR and SSIM measurements of mustGAN, pGAN$_{\mathrm{one}}$ and pGAN$_{\mathrm{many}}$ on the test set. mustGAN outperforms the competing methods in all cases ($p<0.05$), except for SSIM in $\mathrm{T_1}$ and $\mathrm{T_2}$ synthesis ($p>0.05$). On average, mustGAN achieves $0.94\%$ higher SSIM and $1.27$ $dB$ higher PSNR compared to the second-best method (pGAN$_{\mathrm{many}}$). 

\begin{table}[htbp]
\caption{Quality of Synthesis in the IXI Dataset}
\begin{center}
\scalebox{0.72}{
\renewcommand{\arraystretch}{1.4}
\begin{tabular}{lccccccccc}
\hline
&\multicolumn{2}{c}{mustGAN} &\multicolumn{2}{c}{pGAN$_{\mathrm{many}}$} &\multicolumn{2}{c}{pGAN$_{\mathrm{one}}$-A} &\multicolumn{2}{c}{pGAN$_{\mathrm{one}}$-B} \\
\cline{2-9}
& PSNR & SSIM & PSNR & SSIM & PSNR & SSIM & PSNR & SSIM \\
\hline
\multirow{2}{*}{$\mathrm{T_2}$, $\mathrm{PD}$$\rightarrow$$\mathrm{T_1}$}
& $\textbf{29.45}$ & $\textbf{0.947}$ 
& $28.80$ & $0.940$ 
& $28.39$ & $0.934$ 
& $28.73$ & $0.936$ 
\\ 
& $\pm \textbf{1.19}$ & $\pm \textbf{0.012}$ 
& $\pm 1.09 $ & $\pm 0.013$ 
& $\pm 1.17$ & $\pm 0.013$ 
& $\pm 1.18$ & $\pm 0.013$ 
\\
\hline
\multirow{2}{*}{$\mathrm{T_1}$, $\mathrm{PD}$$\rightarrow$$\mathrm{T_2}$} 
& $\textbf{35.89}$ & $\textbf{0.977}$ 
& $34.04$ & $0.964$ 
& $28.52$ & $0.925$ 
& $33.08$ & $0.962$ 
\\ 
& $\pm \textbf{1.20}$ & $\pm \textbf{0.005}$ 
& $\pm 1.18$ & $\pm 0.006$ 
& $\pm 1.18$ & $\pm 0.015$ 
& $\pm 0.99$ & $\pm 0.007$ 
\\
\hline
\multirow{2}{*}{$\mathrm{T_1}$, $\mathrm{T_2}$$\rightarrow$$\mathrm{PD}$} 
& $\textbf{34.40}$ & $\textbf{0.974}$ 
& $33.09$ & $0.967$ 
& $27.80$ & $0.929$ 
& $32.17$ & $0.962$ 
\\ 
& $\pm \textbf{0.97}$ & $\pm \textbf{0.005}$
& $\pm 1.09$ & $\pm 0.005$ 
& $\pm 1.16$ & $\pm 0.012$
& $\pm 1.01$ & $\pm 0.005$ \\
\hline
\end{tabular}
}
\begin{tablenotes}
\item{PSNR and SSIM measurements between the ground truth and synthesized target images from mustGAN, pGAN$_{\mathrm{many}}$ and pGAN$_{\mathrm{one}}$ are given as mean$\pm$std calculated across test subjects for three distinct cases: $\mathrm{T_2}$, $\mathrm{PD}$$\rightarrow$$\mathrm{T_1}$, $\mathrm{T_1}$, $\mathrm{PD}$$\rightarrow$$\mathrm{T_2}$, $\mathrm{T_1}$, $\mathrm{T_2}$$\rightarrow$$\mathrm{PD}$. pGAN$_{\mathrm{one}}$-A receives the $1^{\mathrm{st}}$ source contrast and pGAN$_{\mathrm{one}}$-B receives the $2^{\mathrm{nd}}$ source contrast i.e. $(1,2):(\mathrm{T_2}, \mathrm{PD}),(\mathrm{T_1}, \mathrm{PD}),(\mathrm{T_1}, \mathrm{T_2})$. Boldface marks the model having the highest performance.}.
\end{tablenotes}
\end{center}
\end{table}

\par
Superior performance of mustGAN on the IXI dataset is clearly visible in representative results shown in Fig. 1. Fig. 1a-c respectively display results for $\mathrm{T_1}$-weighted, $\mathrm{T_2}$-weighted and $\mathrm{PD}$-weighted image synthesis. Compared to other methods, mustGAN depicts white-matter tissue with apparently lower noise levels, and gray-matter tissue with sharper tissue boundaries.

\par
Having demonstrated mustGAN on healthy subjects, we next evaluated mustGAN on the ISLES dataset contained images of high/low grade glioma patients. Here we considered: $\mathrm{T_2}, \mathrm{FLAIR}\rightarrow \mathrm{T_1}$; $\mathrm{T_1}, \mathrm{FLAIR}\rightarrow \mathrm{T_2}$; $\mathrm{T_1}, \mathrm{T_2}\rightarrow \mathrm{FLAIR}$. Pairwise comparisons across cross-sections in the test set between the proposed and competing methods are displayed in Fig. 3, Supp. Figs. 7-8 for $\mathrm{T_1}$, $\mathrm{T_2}$ and $\mathrm{FLAIR}$ synthesis, respectively. The enhanced sensitivity of unique source features enables pGAN$_\mathrm{one}$ models to achieve higher PSNR than pGAN$_\mathrm{many}$ in 27.6\% of test samples. Again, mustGAN reduces this proportion to 13.2\%, yielding a substantial improvement over pGAN$_\mathrm{many}$. Table II lists the PSNR and SSIM measurements of mustGAN, pGAN$_{\mathrm{one}}$ and pGAN$_{\mathrm{many}}$ on the test set. mustGAN again outperforms the competing methods in all cases ($p<0.05$), except for SSIM in $\mathrm{T_1}$ synthesis ($p>0.05$). On average, mustGAN achieves $1.29\%$ higher SSIM and $0.84$ $dB$ higher PSNR compared to the second-best method (pGAN$_{\mathrm{many}}$). 

\begin{table}[htbp]
\caption{Quality of Synthesis in the ISLES Dataset}
\begin{center}
\scalebox{0.72}{
\renewcommand{\arraystretch}{1.4}
\begin{tabular}{lccccccccc}
\hline
&\multicolumn{2}{c}{mustGAN} &\multicolumn{2}{c}{pGAN$_{\mathrm{many}}$} &\multicolumn{2}{c}{pGAN$_{\mathrm{one}}$-A} &\multicolumn{2}{c}{pGAN$_{\mathrm{one}}$-B} \\
\cline{2-9}
& PSNR & SSIM & PSNR & SSIM & PSNR & SSIM & PSNR & SSIM \\
\hline
\multirow{2}{*}{$\mathrm{T_2}$, $\mathrm{FLAIR}$$\rightarrow$$\mathrm{T_1}$} 
& $\textbf{28.51}$ & $\textbf{0.929}$ 
& $27.64$ & $0.921$ 
& $25.03$ & $0.886$
& $27.55$ & $0.919$ 
\\
& $\pm \textbf{2.10}$ & $\pm \textbf{0.018}$ 
& $\pm 1.88$ & $\pm 0.017$ 
& $\pm 1.92$ & $\pm 0.015$ 
& $\pm 1.35$ & $\pm 0.015$ 

\\
\hline
\multirow{2}{*}{$\mathrm{T_1}$, $\mathrm{FLAIR}$$\rightarrow$$\mathrm{T_2}$} 
& $\textbf{26.22}$ & $\textbf{0.907}$ 
& $25.55$ & $0.896$
& $24.77$ & $0.883$ 
& $25.19$ & $0.891$ 
\\ 
& $\pm \textbf{1.01}$ & $\pm \textbf{0.014}$ 
& $\pm 0.90$ & $\pm 0.013$ 
& $\pm 0.80$ & $\pm 0.012$ 
& $\pm 0.57$ & $\pm 0.009$ 
\\
\hline
\multirow{2}{*}{$\mathrm{T_1}$, $\mathrm{T_2}$$\rightarrow$$\mathrm{FLAIR}$} 
& $\textbf{26.08}$ & $\textbf{0.910}$ 
& $25.11$ & $0.894$
& $24.91$ & $0.889$ 
& $23.32$ & $0.861$ 
\\ 
& $\pm \textbf{1.04}$ & $\pm \textbf{0.016}$ 
& $\pm 0.81$ & $\pm 0.013$
& $\pm 0.94$ & $\pm 0.015$
& $\pm 0.67$ & $\pm 0.012$
\\
\hline
\end{tabular}
}
\begin{tablenotes}
\item{PSNR and SSIM measurements between the ground truth and synthesized target images from mustGAN, pGAN$_{\mathrm{many}}$ and pGAN$_{\mathrm{one}}$ are given as mean$\pm$std calculated across test subjects for three distinct cases: $\mathrm{T_2}$, $\mathrm{FLAIR}$$\rightarrow$$\mathrm{T_1}$, $\mathrm{T_1}$, $\mathrm{FLAIR}$$\rightarrow$$\mathrm{T_2}$, $\mathrm{T_1}$, $\mathrm{T_2}$$\rightarrow$$\mathrm{FLAIR}$. pGAN$_{\mathrm{one}}$-A receives the $1^{\mathrm{st}}$ source contrast and pGAN$_{\mathrm{one}}$-B receives the $2^{\mathrm{nd}}$ source contrast i.e. $(1,2):(\mathrm{T_2}, \mathrm{FLAIR}),(\mathrm{T_1}, \mathrm{FLAIR}),(\mathrm{T_1}, \mathrm{T_2})$. Boldface marks the model having the highest performance.}.
\end{tablenotes}
\end{center}
\end{table}
\par
Superior performance of mustGAN on the ISLES dataset is also clearly visible in the representative results shown in Fig. 2. Fig.2a-c respectively display results for $\mathrm{T_1}$-weighted, $\mathrm{T_2}$-weighted and $\mathrm{FLAIR}$ image synthesis. Compared to other methods, mustGAN depicts white-matter tissue with apparently lower noise levels, and gray-matter tissue with sharper tissue boundaries.

\begin{figure}[htbp]
\centerline{\includegraphics[width=1\linewidth]{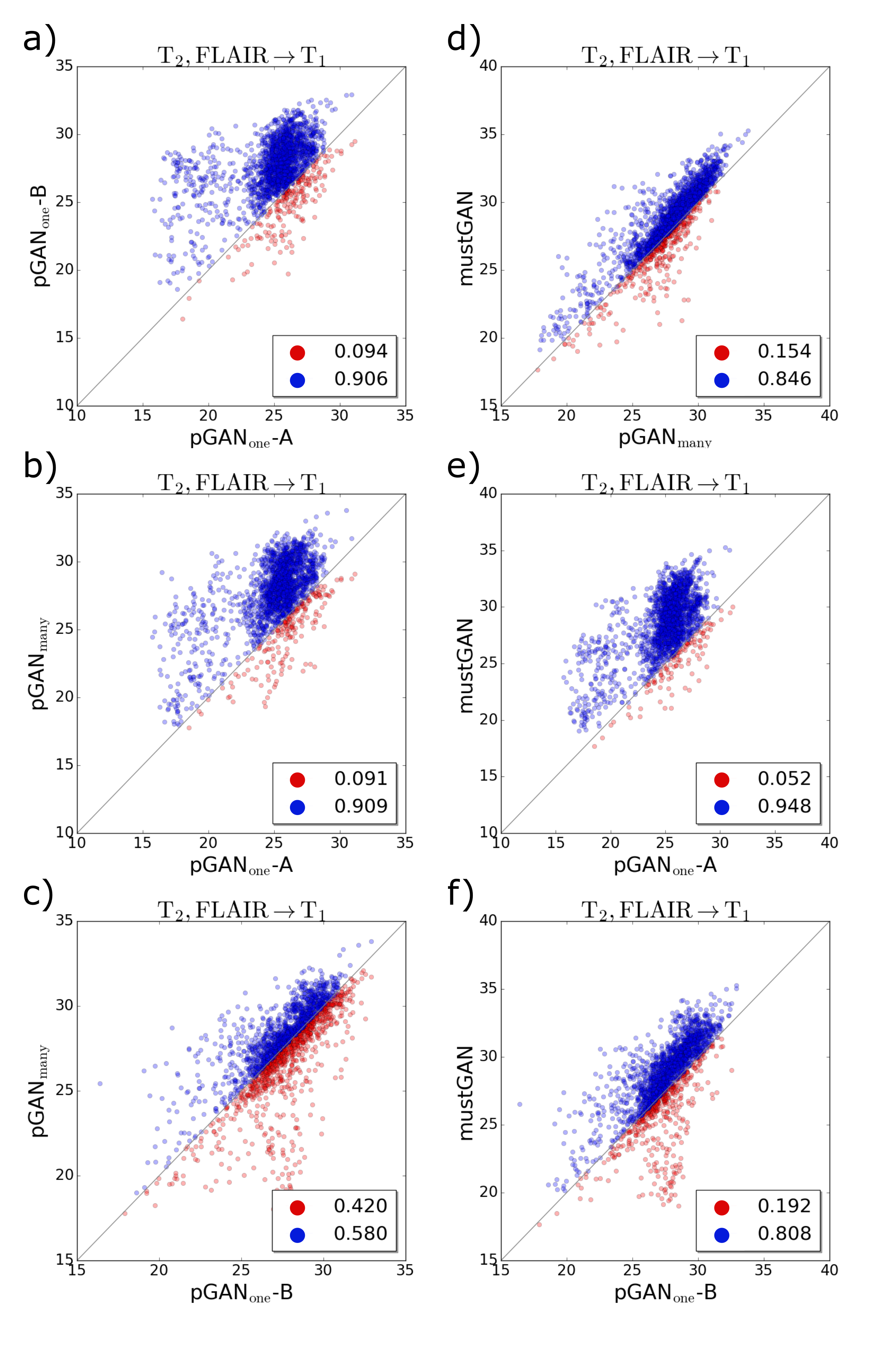}}
\caption{Methods were compared in terms of quality of $\mathrm{T_1}$ synthesis in the ISLES dataset: a) pGAN$_{\mathrm{one}}$-A versus pGAN$_{\mathrm{one}}$-B, b) pGAN$_{\mathrm{one}}$-A versus pGAN$_{\mathrm{many}}$, c) pGAN$_{\mathrm{one}}$-B versus pGAN$_{\mathrm{many}}$, d) pGAN$_{\mathrm{many}}$ versus mustGAN, e) pGAN$_{\mathrm{one}}$-A versus mustGAN, f) pGAN$_{\mathrm{one}}$-B versus mustGAN. Note that pGAN$_\mathrm{one}$-A receives $\mathrm{T_2}$-weighted images as input and pGAN$_\mathrm{one}$-B receives $\mathrm{FLAIR}$ images as input. Scatter plots show PSNR measurements for methods under comparison, and each point denotes a cross-section in the test set. The proportion of test samples in which either method yields superior performance is also noted in figure legends (blue font for the method on the vertical axis, red font for the method on the horizontal axis).}
\label{fig}
\end{figure}

\section{Discussion}
A within-modality synthesis method was introduced for multi-contrast MRI based on conditional generative adversarial networks. The proposed method aggregates information across one-to-one streams that are sensitive to unique information in individual source contrasts and a many-to-one stream that is sensitive to shared information across multiple source contrasts. Enhanced synthesis performance was demonstrated in a number of synthesis tasks on brain MRI datasets from normals and glioma patients. Compared to isolated one-to-one or many-to-one methods, mustGAN recovered higher quality images with reduced noise and improved sharpness. 
\par 
A prior state-of-the-art method for multi-contrast MRI synthesis, Multimodal, is based on an encoder-decoder architecture with standard convolutional layers \cite{multimodal}. Given multiple source contrasts, Multimodal learns contrast-invariant latent representations for source images by enforcing latent representations from separate encoders to be similar. These individual latent representations are then fused across source contrasts via a maximum function, and the decoder recovers target images based on fused representations. For improved sensitivity to unique features of individual sources, mustGAN does not explicitly seek similarity across latent representations in one-to-one streams and instead uses a separate many-to-one stream to capture shared representations across source contrasts. While the position of the fusion block is fixed to the initial layer of the decoder in Multimodal, the proposed method adaptively modifies the position of the fusion block to optimize the task-specific performance. Moreover, unlike Multimodal that uses a mean absolute error metric, mustGAN uses adversarial loss that has been demonstrated to better capture high-spatial-frequency information \cite{pgan_cgan}.  
\par
Several recent studies have proposed GAN-based architectures for multi-contrast MRI synthesis. In \cite{pgan_cgan}, we have proposed pGAN that uses a conditional GAN models for one-to-one synthesis. In \cite{mmgan}, a many-to-one generalization of pGAN was proposed, MM-GAN, that receives as input multiple source contrasts for enhanced synthesis performance. Note that MM-GAN fuses multiple source contrast at the input level by treating them as separate information channels, and so it is similar in nature to pGAN$_{\mathrm{many}}$ implemented here. Our results indicate that, compared to pGAN$_{\mathrm{many}}$, mustGAN achieves enhanced sensitivity to unique features of individual source contrasts due to the presence of additional one-to-one streams.
\par
An important requirement for succesful training of deep network architectures is the availability of large datasets. The current implementation of mustGAN assumes availability of paired source-target images from the same group of subjects. However, size of paired datasets might be limited especially when relative less common contrasts are involved. In such cases, several lines of improvement can be considered. (1) When the source images are all paired but the target images are unpaired, the pixel-wise loss used in one-to-one and many-to-one streams can be replaced with a cycle-consistency loss. Training procedures for the cycle-consistent models can be adopted from prior studies for both one-to-one \cite{pgan_cgan} and many-to-one \cite{collagan} GAN models. (2) When the source images are also unpaired, the many-to-one stream can be removed. The one-to-one streams can again be trained with a cycle-consistency loss and then fused for enhanced performance. 
\par
The proposed network model takes as input spatially registered source and target images. The datasets analyzed in this study were either pre-registered, or registration was implemented as a pre-processing step (see Methods for procedures on the IXI dataset). When an end-to-end network alternative is desired, deep-network-based registration models \cite{registration} can instead be cascaded to the input of mustGAN to spatially align source-target images. It remains important future work to investigate potential benefits of an end-to-end registration approach over pre-processing.

\bibliography{ref} 
\bibliographystyle{ieeetr}

\vspace{12pt}

\end{document}